\documentclass[preprint,sort&compress]{elsarticle}

\newcommand{\cm}{cm$^{-1}$ }
\newcommand{\CM}{cm$^{-1}$}
\newcommand{\mm}{$\mu$m}

\date{\today}
\title{IRFEL Selective Irradiation of Amorphous Solid Water: from Dangling to Bulk Modes}

\author[1]{S. Coussan}
\author[1,2]{J.~A. Noble}
\author[3]{H.~M.~Cuppen}
\author[4]{B. Redlich}
\author[5]{S. Ioppolo}

\address[1]{Aix-Marseille Univ, CNRS, PIIM, Marseille, France.}
\address[2]{School of Physical Sciences, University of Kent, Canterbury CT2 7NH, U.K.}
\address[3]{Radboud University, Institute for Molecules and Materials, Nijmegen 6525 AJ, The Netherlands.}
\address[4]{Radboud University, Institute for Molecules and Materials, FELIX Laboratory, Nijmegen 6525 ED, The Netherlands.}
\address[5]{School of Electronic Engineering and Computer Science, Queen Mary University of London, Mile End Road, London E1 4NS, United Kingdom.}

\begin{document}

\begin{abstract}
Amorphous solid water (ASW) is one of the most widely studied solid phase systems. A better understanding of the nature of inter- and intra-molecular forces in ASW is, however, still required to correctly interpret the catalytic role of ASW in the formation and preservation of molecular species in environments such as the icy surfaces of Solar System objects, on interstellar icy dust grains and potentially even in the upper layers of the Earth's atmosphere. In this work, we have systematically exposed porous ASW (pASW) to mid-infrared radiation generated by a free-electron laser at the HFML-FELIX facility in the Netherlands to study the effect of vibrational energy injection into the surface and bulk modes of pASW. During multiple sequential irradiations on the same ice spot, we observed selective effects both at the surface and in the bulk of the ice. Although the density of states in pASW should allow for a fast vibrational relaxation through the H-bonded network,  part of the injected energy is converted into structural ice changes as illustrated by the observation of spectral modifications when performing Fourier transform infrared spectroscopy in reflection-absorption mode. Future studies will include the quantification of such effects by systematically investigating ice thickness, ice morphology, and ice composition.

\end{abstract}

\maketitle

\section{Introduction}
\begin{figure}[htb]
\centering
\includegraphics[width=\textwidth]{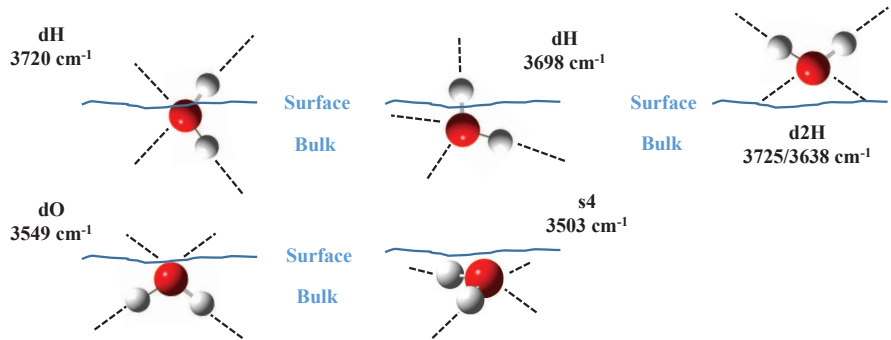}
\caption{Schematic of the five possible orientations of water molecules at the surface of ASW, known as ``dangling'' modes.}\label{fig:Danglings}
\end{figure}

Amorphous solid water (ASW) is one of the most widely studied solid phase systems in physical chemistry. It is ubiquitously present in the interstellar medium (ISM), in certain environments in the Solar system, and has been postulated to exist for short periods in the Earth's upper mesosphere ($\sim$90~km)\cite{Jenniskens95,van Dishoeck21,Newman08,Murray10}. At its surface, ASW presents characteristic bonds known as `dangling' modes that are electronic doublets and/or OH bonds that allow small molecules such as O$_2$, N$_2$, CO, H$_2$O, and CH$_3$OH to bind and accrete. ASW can therefore act as a catalyst, facilitating thermally- and/or photochemically-initiated surface reactions that can potentially lead to the formation of biologically relevant species such as amino acids\cite{van Dishoeck13,Ioppolo21}. Both the surface and bulk of ASW can therefore be key to the formation and accumulation of other species in a variety of environments. In porous ASW (pASW), the surface includes the surfaces of internal pores. Recent studies\cite{Noble14A,Noble14B,Coussan15} showed that pASW and, particularly, the absence of long-range order in its molecular arrangement, can be studied by selectively injecting vibrational energy resonantly into its OH stretching absorption modes in the infrared (IR) spectral range. In these studies, some of us reported the effects of selective IR irradiations on the dangling modes of pASW. We expect, due to a fast relaxation of the injected vibrational energy through the H-bonded network, no irreversible change in the ice spectrum. However, in our prior studies we showed that, upon selective IR irradiations by means of an optical parametric oscillator (OPO) laser, the four types of dangling bond (doubly and triply-bound dH, dO and s4; Figure~\ref{fig:Danglings}) are able to isomerize toward a different configuration, which we called d2H, consisting of a water monomer interacting with the ASW surface through its two electronic doublets\cite{Noble14A,Noble14B,Coussan15}. We also showed that the two dH bands, located at 3720 and 3698~\cm, respectively, are inhomogeneously broadened. We then suggested that this inhomogeneity is due to a distribution of classes of oscillators at the surface \textit{i.e.} dangling bonds at a variety of angles.  Our work suggested that the injection of vibrational energy into ASW through other selective irradiations, \textit{i.e.} via the combination, bending, and libration modes, should be investigated, which is the subject of the present study. To date, little is know on the topic.

Adopting a similar approach, Krasnopoler and George\cite{Krasnopoler98} explored the desorption effects induced by infrared free-electron laser (IRFEL) selective irradiation of the cubic crystalline ice Ic deposited at 140~K and maintained at 110~K under ultrahigh vacuum conditions. They irradiated the ice between 2.8 and 3.4~$\mu$m ($\approx$ 3570 and 2940~\CM). The laser was pulsed at 1 Hz (with a shutter), and they detected desorbed species with a quadrupole mass spectrometer. Their main findings are hereby listed: \textit{(i)} the desorption peak was observed at 3.0~\mm ($\approx 3330$~\CM); \textit{(ii)}  the resonant desorption spectrum was however not identical to the infrared absorption spectrum of the ice as much higher water desorption signals were obtained than expected at IRFEL wavelengths shorter than the peak of the infrared absorption spectrum at 3.1~\mm; \textit{(iii)} this shift of the resonant desorption spectrum to shorter wavelengths was explained by the melting of the ice film and the temperature-dependent absorption coefficient of the O-H stretching vibration; \textit{(iv)} the authors noticed that desorption yield and beam penetration were strongly coupled to irradiation frequency and laser repetition rate as well as structure and temperature of the ice. Hence, this work suggested selectivity effects of on-resonance IR irradiation of a crystalline ice sample. A different group worked on the IR irradiation of crystalline ice samples a few micrometers thick and a few millimeters thick prepared \emph{in~situ} and \emph{ex~situ}, respectively \cite{Mihisan05,Focsa03}. These works mainly focused on the detection of induced desorption and formation of large water clusters upon the exposure of an ice to a LiNbO$_3$ tunable IR laser. Using a `table-top' tunable IR OPO laser pulsed at 1~Hz, Focsa \textit{et al.}\cite{Focsa03} repeated similiar experiments to those presented by Krasnopoler and George\cite{Krasnopoler98}. They too observed a direct link between the desorption yield, the irradiation frequency, the laser power and its repetition rate. Focsa \textit{et al.}\cite{Focsa03} irradiated Ic maintained at 100~K  and observed a desorption peak located at 3225~\CM. Mihisan \textit{et al.}\cite{Mihisan05} irradiated an ice maintained at 90~K at 3.1~$\mu$m ($\approx$ 3225~\CM; laser pulsed at 10~Hz) and detected the formation of aggregates of H$_3$O$^+$(H$_2$O)$_n$, where n $\approx$ 100. They noticed a two-step desorption process: a phase explosion regime followed by vaporization, leading to what they called ``small'' and ``large'' clusters, respectively.

The take home-message of these studies is that there is strong evidence for IR selective irradiation of ordered water ices to be an appropriate tool to study its global structure. Our early studies on the pASW surface\cite{Noble14A,Noble14B,Coussan15} started the investigation of surface reorganization upon laser exposure. Such results were confirmed by our more recent work on IRFEL irradiation of $\sim$ 0.25~$\mu$m thick pASW and crystalline Ic ices in the MIR spectral range\cite{Noble20}. This combined laboratory and theoretical work showed that thin samples of pASW can absorb vibrational energy in the MIR spectral range causing local energy transfer effects within the ice and at the surface that leads to a reorganization of the ice toward a more ordered material \cite{Noble20}. In agreement with previous work\cite{Krasnopoler98,Focsa03}, IRFEL irradiation of Ic mostly induced desorption of the ice. Here we report on a series of selective irradiations of $\sim$ 0.9~$\mu$m thick pASW carried out with the IRFEL laser at the HFML-FELIX facility in the Netherlands to provoke and allow the further investigation of ice changes. Building on the previously mentioned literature work, these changes are discussed on the basis of photoinduced ice modifications and/or thermal rearrangements of the ice occurring upon vibrational energy dissipation.

\section{Experimental}
Experiments were carried out using the laboratory ice surface astrophysics (LISA) ultrahigh vacuum (UHV) end-station at the HFML-FELIX facility, Radboud University, The Netherlands. Details of the experimental setup configuration used in this work are given in Noble \textit{et al.}\cite{Noble20}. Here, we discuss the experimental methodology applied during a standard experiment. Samples of pASW were prepared using deionised water subjected to multiple freeze-pump-thaw cycles under vacuum to remove dissolved gases. Ices were produced by background deposition on the four faces of a gold-plated oxygen-free high thermal conductivity (OFHC) copper cube maintained at 18~K by a 204SB (Sumitomo) cryogenic head. The substrate temperature is controlled in the range 15-300~K by means of a model 335 cryogenic temperature controller (LakeShore) that reads temperature through an uncalibrated DT-670B-SD silicon diode (with $\pm0.5$~K tolerance and $\pm0.01$~K accuracy) mounted with a CO adapter at the bottom of the gold-plated copper cube and regulates a Kapton tape heater (Omega) connected above the OFHC copper block. Samples were prepared 24h before the irradiations in order to obtain thermalized ices. The residual pressure in the main chamber was better than $5\times10^{-8}$~mbar. IR spectra were recorded in reflection mode at an incident angle of 18$^\circ$ by using a Fourier transform infrared (FTIR) spectrometer (Bruker 80v) equipped with a doped Ge KBr beamsplitter and an external mercury cadmium telluride (MCT) detector (5000-600~\CM). The internal spectrometer vacuum was better than 10$^{-3}$~mbar and the IR line between the spectrometer and the detector was purged with pure nitrogen. FTIR spectra were acquired with 256 co-added scans and at a resolution of 0.5~\cm. The molecular column density $C$ (molecules cm$^{-2}$) of the deposited ices at 18~K is calculated by integrating a modified Beer-Lambert Equation (Eq. 1) over the OH stretching mode vibration band of ASW:

\begin{equation}
 C = \ln(10) \frac{\int{\tau(\nu) d\nu}}{A'(\nu)},
\end{equation}
where $A'(\nu)$ is the integrated band strength and $\tau(\nu)$ is the optical depth (cm$^{-1}$). Based on the work of Gerakines \textit{et al.}\cite{Gerakines95}, we chose $A'(\nu)~=~2.0\times10^{-16}$~cm~molecule$^{-1}$ as the integrated band strength for pure pASW. During deposition of all investigated ices, all parameters are kept the same, \emph{e.g.} deposition time ($\sim$1080~s) and residual pressure in the main chamber ($1\times10^{-6}$~mbar). Once calculated, the column density is used to determine the ice thicknesses in $\mu$m as per Eq. 2:

\begin{equation}
 d = \frac{C Z}{\rho N_A} \times 10^4,
\end{equation}
where $Z$ is the molecular mass of water (g~mol$^{-1}$), $\rho$ is the density of the ice, which we have taken to be 0.94~g~cm$^{-3}$ \cite{Jenniskens94}, and $N_A$ is the Avogadro constant ($6.02\times10^{23}$~molecule~mol$^{-1}$). Hence, we estimate pASW to be $\sim$0.9~$\mu$m thick with a deposition rate of $\sim$0.8~nm~s$^{-1}$.

IR photons at an incident angle of 54$^\circ$ with respect to the gold-plated flat surface were provided by the free electron laser FELIX-2 (3-45~$\mu$m). For the FEL beam settings chosen in this work, FELIX-2 consisted of 10 $\mu$s macropulses at 5 (or 10) Hz repetition rate, containing a train of 10$^4$ micropulses of $\sim$ 5~ps, which are 1~ns apart. The average laser power in the $\nu_{OH}$ region (3800-3000 \CM) was 73~mW, and reached 835~mW at 11.8~\mm (847~\CM, libration mode) for a full width at half maximum (FWHM) comprised between 0.2 and a few $\%$ of $\delta\lambda/\lambda$. The typical duration of one irradiation at one frequency was 15~min to ensure the saturation of effects. The laser spot was $\approx$ 2~mm diameter while that of the FTIR spectrometer was 3~mm, allowing the full irradiated area to be probed. Since the FTIR beam was larger than the FEL beam, a part of ice probed by the FTIR was not exposed to IRFEL irradiation. Hence, we decided to investigate FTIR difference spectra acquired before and after IRFEL irradiation to highlight changes in the ice. Series of irradiations were carried out from ``high'' to ``low'' and from ``low'' to ``high'' wavenumbers (hereafter referred to from ``blue to red'' frequencies (from 2.7 to 3.3~\mm) and from ``red to blue'' frequencies (from 3.3 to 2.7~\mm), respectively) across the water OH stretching mode. \emph{Stephane: an element of response to the referee's question is that FELIX is in microns due to its low precision and we analyse the data in wavenumbers due to the need for higher precision. This explains the shifting units used throughout the text.} Unirradiated ices were also exposed to single IRFEL irradiations at the combination mode at 4.5~$\mu$m (2222~\CM), the bending mode at 6.0~$\mu$m (1666~\CM), and the libration mode at 11.7~$\mu$m (847~\CM). The possibility of adjusting the sample height allowed us to start new irradiation series on other unirradiated ice spots obtained during the same single ice deposition.

\section{Experimental Results and Discussion}

\begin{figure}[htb]
\centering
\includegraphics[width=\textwidth]{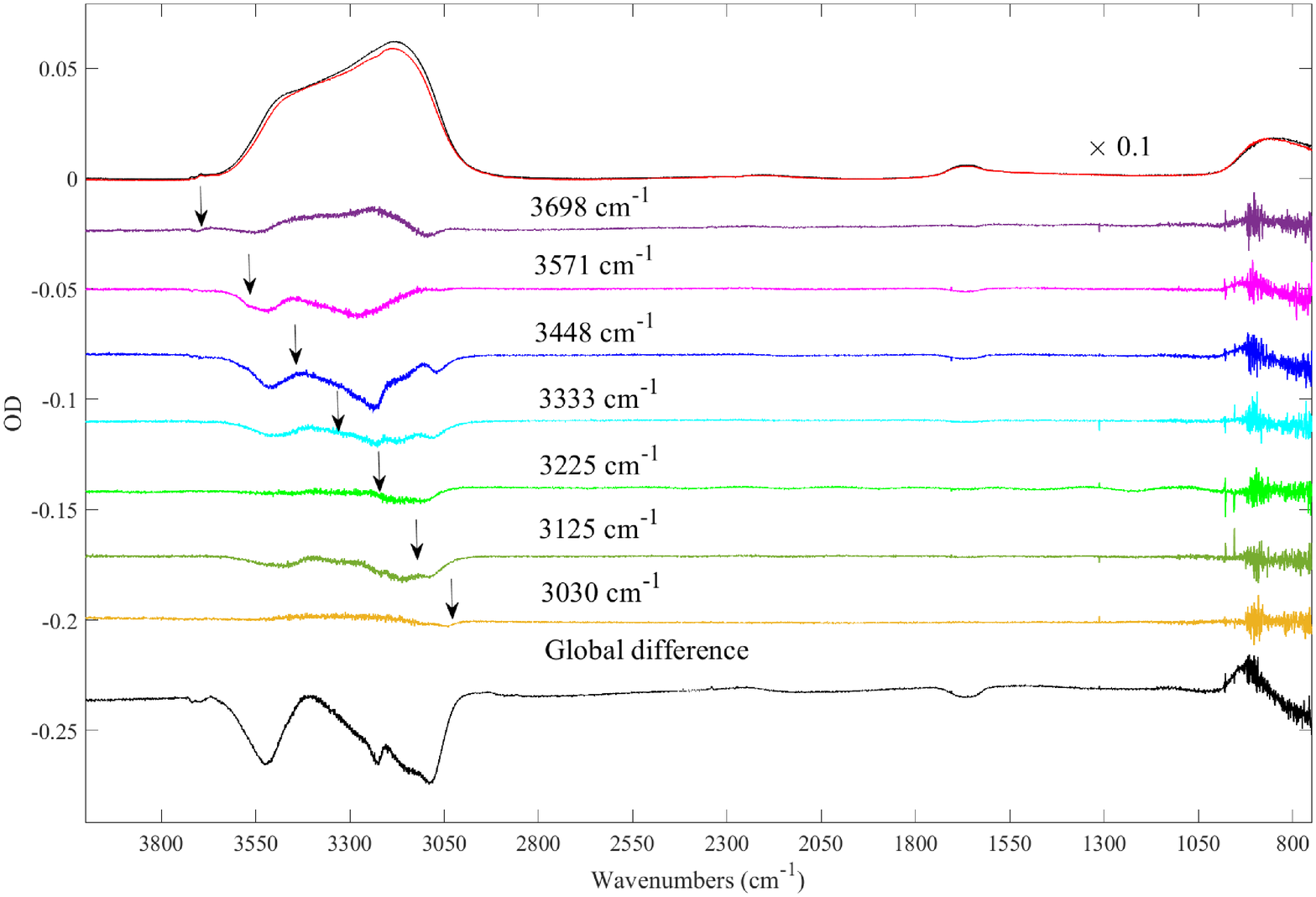}
\caption{Series of selective IRFEL irradiations at 10~Hz laser repetition rate, from ``blue to red'' frequencies across the 3~$\mu$m OH stretching mode vibration of water, \emph{i.e.} from dangling to bulk modes of a pASW sample held at 18~K on the same ice spot. From top to bottom, black: FTIR spectrum acquired after deposition; red: FTIR spectrum acquired after the whole series of irradiations; purple: difference spectrum acquired before and after an IRFEL irradiation at 3698~\CM with a 60~min duration and at 52~mW average power; magenta: difference spectrum acquired before and after irradiation at 3571~\CM with 15~min duration and 73~mW average power; blue: difference spectrum acquired before and after irradiation at 3448~\CM with 15~min duration and 85~mW average power; light blue: difference spectrum acquired before and after irradiation at 3333~\CM with 15~min duration and 95~mW average power; light green: difference spectrum acquired before and after irradiation at 3225~\CM with 15~min duration and 87~mW average power; green: difference spectrum acquired before and after irradiation at 3125~\CM with 15~min duration and 85~mW average power; orange: difference spectrum acquired before and after irradiation at 3030~\CM with 15~min duration and 75~mW average power; black: global difference spectrum acquired before any IRFEL irradiation and after the whole series of irradiations on the same ice spot. Spectra are offset for clarity.}\label{fig:BulkIR}
\end{figure}

Figure~\ref{fig:BulkIR} presents spectra from the ``blue to red'' series of successive irradiations on a single spot of a $\sim$ 0.9~$\mu$m thick pASW sample, with irradiations from 3698 to 3030~\CM, covering the surface and bulk OH stretching modes. Apart from the upper two FTIR spectra of the ice acquired before and after all IRFEL irradiations (black and red lines, respectively), Figure~\ref{fig:BulkIR} presents irradiation results as difference spectra, which are calculated as follows: [(spectrum after irradiation)-(spectrum before irradiation)].  If one considers uniquely the global difference spectrum (lower trace, in black), it appears that, apart from at $\sim$ 930~\CM, where the signal-to-noise is very low, only overall decreases in spectral intensity are observed, \textit{i.e.} negative peaks.
Hence, the question as to whether such losses are due to induced photodesorption only, \textit{i.e.} loss of material, should be raised. There is some prior experimental evidence that irradiating at the surface oscillator modes of pASW does not lead to molecular desorption\cite{Noble14A,Noble14B,Coussan15} (at least within the limit of detection of the mass spectrometer), while irradiations at the bulk modes of Ic does induce desorption\cite{Krasnopoler98,Mihisan05,Focsa03,Noble20}. In our recent study of the IRFEL irradiation of pASW\cite{Noble20} we could explain all modifications observed in the FTIR spectrum of a thinner (0.25~\mm) pASW sample by structural changes in the ice. The question of structural and thickness dependence will be addressed more fully in a forthcoming article. However, although there is evidence for photoinduced sublimation, this is certainly not the only process at play here upon IR selective irradiation of the ice. \emph{Stephane: we are specifically going to talk about the role of desorption in thicker ices in a future paper so we must be consistent here. Photodesorption is an important process and can't be minimised.} Firstly, unlike the case of Ic\cite{Noble20}, we do not observe a homogeneous decrease over the whole bulk band of pASW upon IRFEL exposure. Pure desorption would have induced a global `homogeneous' decrease in all oscillators and not the observed irradiation-dependent appearance of specific negative peaks, as seen in Figure~\ref{fig:BulkIR} in the difference spectra. To a first approximation, and considering the two dH bands as one class, we count six different classes of oscillators involved in spectral changes upon irradiation. Based on our past work on dangling modes\cite{Noble14A,Noble14B,Coussan15}, it is possible to explain the effect of the first irradiation at 3698~\CM, \textit{i.e.} on the dH dangling bonds (purple solid line in Figure~\ref{fig:BulkIR}). We observe the rearrangement of those modes, inducing decreases at around 3709, 3554, and 3095~\CM, \textit{i.e.} the average of the two dH stretching mode frequencies, a frequency intermediate between that of the dO and s4 frequencies, and --potentially-- a frequency corresponding to their first bending mode harmonic ($2\nu_2$), respectively. A decrease around 1649~\CM, in the $\nu_2$ bending region, is also detected. The two increases, observed at about 3415 and 3243~\CM, are due to bulk oscillators. Differences between dH irradiations with an LiNbO$_3$ OPO IR laser\cite{Noble14A,Noble14B,Coussan15} and the IRFEL can be explained if we consider the FWHM of each laser. In the case of the OPO, its FWHM is only 1.5~\CM, leading to a more `selective' irradiation than that of FELIX-2, which has, at best, a FWHM of $\approx$ 10~\CM at these frequencies. Moreover, the temporal structure of the two lasers, the size of their spots at the target, their power at the same frequency, and the way we deposited the ices should be also taken into account. Krasnopoler and George\cite{Krasnopoler98} observed Ic surface melting effects upon IR irradiation at the same frequencies. However, Noble \textit{et al.}\cite{Noble20} showed that there are clear differences between IRFEL irradiations of pASW and Ic. Therefore, we cannot discard the possibility of a photo-induced mechanism involving breaking and creation of H-bonds during vibrational energy redistribution.

\emph{Stephane: we propose to add a table here to state all frequencies and classes (both in the unirradiated ice and in the irradiated ice) in order to accommodate the point of the referee. This would be in addition to figure 1, since this only includes surface modes.}

The next irradiation in the ``blue to red'' series was performed at 3571~\cm (magenta solid line in Figure~\ref{fig:BulkIR}), which is at the base of the blue wing of the bulk OH stretch and overlaps with the dO (3549~\CM) and s4 (3503~\CM) surface modes. Weak decreases are observed at the dH positions, while the most intense decreases are centered at $\approx$ 3525 and 3284~\CM with the 1663~\cm bending band decreasing as well. The effects of the irradiation at 3448~\cm (blue solid line in Figure~\ref{fig:BulkIR}) are similar but include a more intense decrease at 3070~\CM. The only observed increase upon the latter two irradiations is at $\approx$ 930~\CM, in the libration mode range, but again the signal-to-noise is very low in this spectral region. Subsequent irradiations lead to the same observation, \textit{i.e.} not a hole-burning phenomenon at the precise frequency of irradiation, but rather a systematic decrease of the same bands, albeit at lower intensity with respect to the initial irradiations.
It should be noted here that during irradiation of bulk oscillators at 2.9, 3.0 and 3.1~$\mu$m (3448, 3333, and 3225~\CM), we observed reproducible temperature increases of 0.03, 0.06 and 0.07~K, respectively. This is the sign that ice is thermally excited upon irradiation at its strongest absorptions, that is at frequencies where the most OH bulk oscillators were excited.
The temperature dropped as soon as irradiation was stopped.  Moreover, the further on in the ``blue to red'' irradiation series we reach in terms of number of irradiations, the fewer FTIR spectral changes are observed in the ice, in line with the fact that sequential IRFEL irradiations at the same ice spot saturate the observed effects (\textit{i.e.} only a small fraction of the oscillators that make up the ice surface and bulk within the irradiated volume are capable of rearranging following vibrational excitation under IR irradiation. In the case of a series of irradiations on the same ice, these effects typically decrease from one irradiation to the next).

\begin{figure}[htb]
\centering
\includegraphics[width=\textwidth]{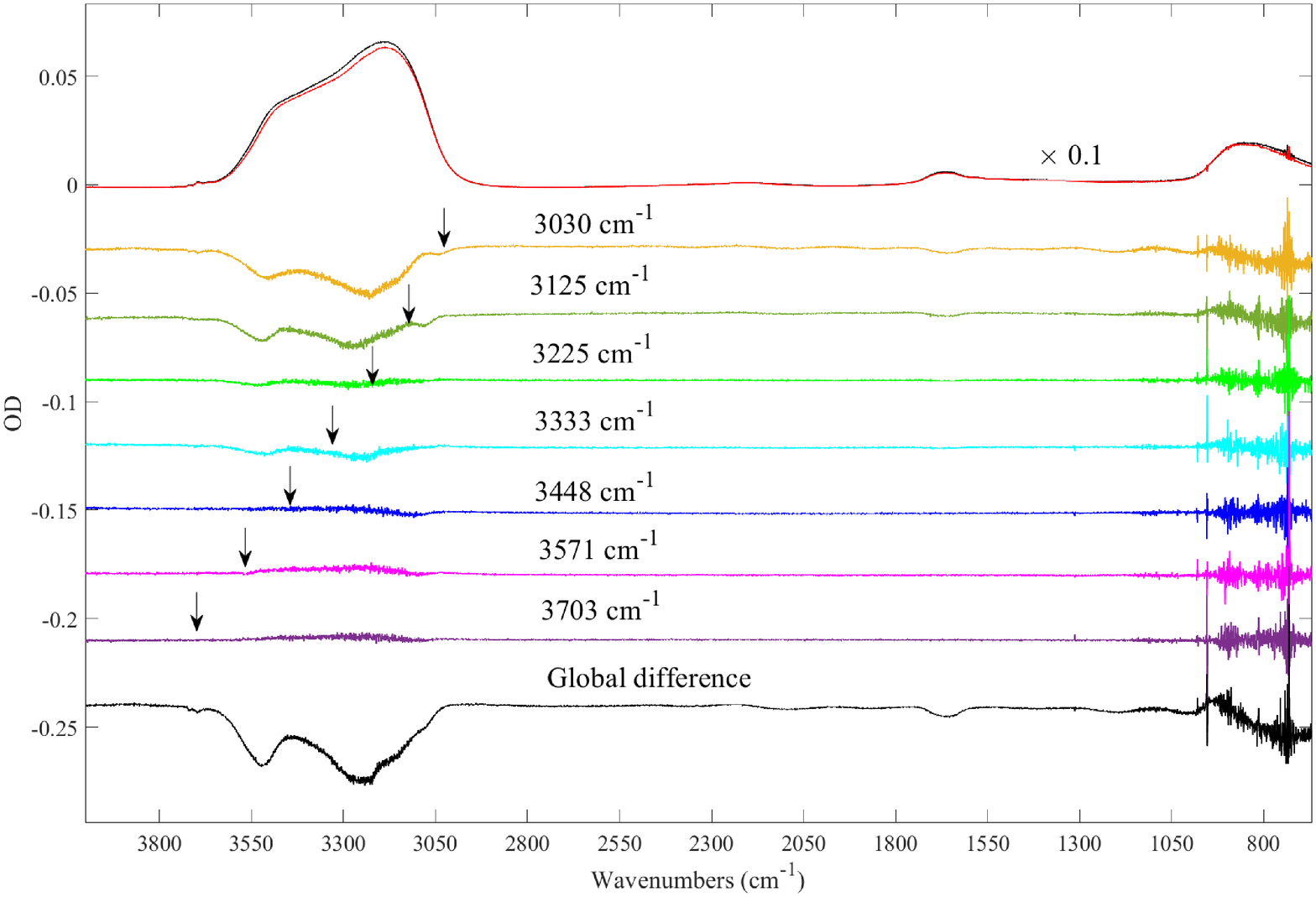}
\caption{Series of selective IRFEL irradiations at 5~Hz laser repetition rate, from ``red to blue'' frequencies across the 3~$\mu$m OH stretching mode vibration of water, \emph{i.e.} from bulk to dangling modes of a pASW sample held at 18~K on the same ice spot. From top to bottom, black: FTIR spectrum acquired after deposition; red: FTIR spectrum acquired after the whole series of irradiations; orange: difference spectrum acquired before and after irradiation at 3030~\CM with 15~min duration and 58~mW average power; green: difference spectrum acquired before and after irradiation at 3125~\CM with 15~min duration and 55~mW average power; light green: difference spectrum acquired before and after irradiation at 3225~\CM with 15~min duration and 40~mW average power; light blue: difference spectrum acquired before and after irradiation at 3333~\CM with 15~min duration and 48~mW average power; blue: difference spectrum acquired before and after irradiation at 3448~\CM with 15~min duration and 43~mW average power; magenta: difference spectrum acquired before and after irradiation at 3571~\CM with 15~min duration and 36~mW average power; purple: difference spectrum acquired before and after IRFEL irradiation at 3703~\CM with 15~min duration and 50~mW average power; black: global difference spectrum acquired before any IRFEL irradiation and after the whole series of irradiations on the same ice spot. Spectra are offset for clarity.}\label{fig:BulkIR2}
\end{figure}

Figure~\ref{fig:BulkIR2} shows spectra from the series of successive irradiations from ``red to blue" frequencies on a single spot of a $\sim$ 0.9~$\mu$m thick pASW sample, with irradiations from 3030 to 3703~\CM. A comparison between the global difference spectra (lower black lines) in Figures~\ref{fig:BulkIR} and \ref{fig:BulkIR2} suggests that the overall changes in the ice are qualitatively very similar regardless of whether the IRFEL irradiation series is performed from blue to red or from red to blue. Considering that we did not inject energy into the ice the same way each time, that is we moved from surface to bulk in the first series (Figure \ref{fig:BulkIR}) and from bulk to surface in the second series (Figure \ref{fig:BulkIR2}) the global effect is essentially identical. Of course, the categorisation of surface versus bulk irradiation is a simplification, since, for example, irradiation at the frequency of the 2$\nu_2$ vibrational mode could potentially excite both surface and bulk oscillators, but it falls in the frequency range that is dominated by the OH stretching mode of bulk water molecules and therefore we categorize it as ``bulk''. Regardless of the irradiation order, the final spectrum is the same, in spite of the fact that each stepwise sequential irradiation effect is different. For each individual irradiation frequency, larger changes are obtained if the frequency is closer to the beginning of the irradiation sequence  (\textit{e.g.} larger effect for the 3030~\cm irradiation in Figure~\ref{fig:BulkIR2} (first irradiation) than for the irradiation at the same frequency in Figure~\ref{fig:BulkIR} (final irradiation)). Therefore, selective IRFEL irradiations excite and impact specific classes of molecules belonging to certain orientations and relative positions in the ice surface and bulk and/or classes of oscillators. Such effects are frequency and irradiation history dependent. Hence, these changes do not occur due to a heating effect described by Boltzmann statistics.
We can also divide all the IRFEL irradiations presented here into two main categories: \textit{(i)} irradiations on dangling modes that induce a rearrangement of those modes causing a decrease in surface mode intensities with a corresponding increase in bulk mode intensities (\textit{e.g.} irradiations at 3698~\cm in Figure \ref{fig:BulkIR}, and at 3571 and 3703~\cm in Figure~\ref{fig:BulkIR2}); and \textit{(ii)} irradiations on bulk modes that affect both bulk and surface. As expected, if the bulk re-organizes, then the ice surface also changes and vice-versa. During irradiation of the bulk modes, we observe almost uniquely decreases with the exception of the $\approx$ 930~\cm libration band, which seems to increase, although the signal-to-noise ratio is low at these frequencies (see larger level of noise below 1000~\cm in the difference spectra of Figures~\ref{fig:BulkIR} and \ref{fig:BulkIR2}). The positive contribution around the libration mode is in agreement with a band peak position shift toward higher frequencies, as observed in FTIR spectra of water ice samples either deposited at or annealed to higher temperatures, suggesting a restructuring of the ice toward a more ordered form.

\begin{figure}[htb]
\centering
\includegraphics[width=\textwidth]{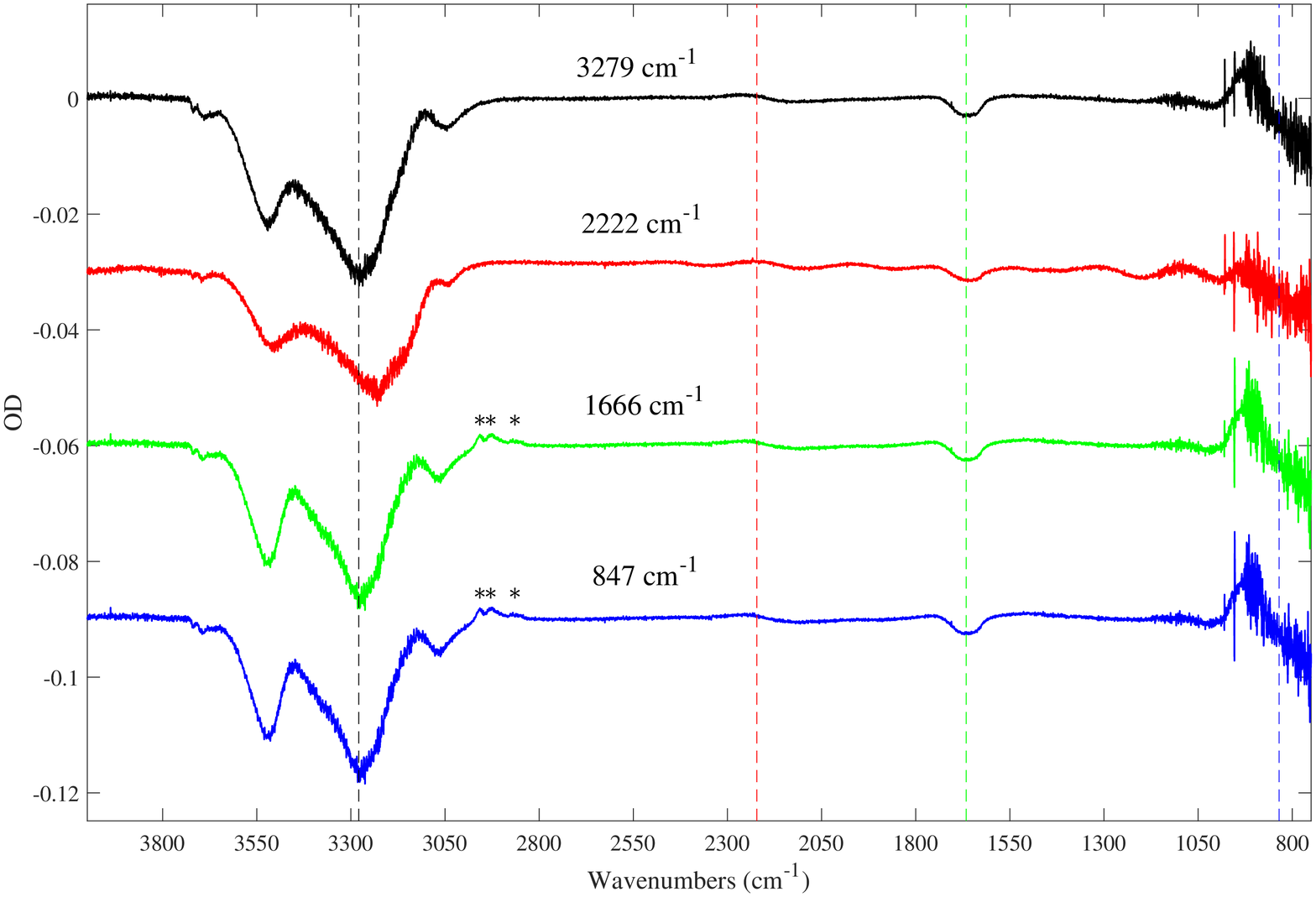}
\caption{Selective IRFEL irradiations of native, unirradiated pASW ices at 18~K with a 5~Hz laser repetition rate. Irradiation frequencies are reported above the respective FTIR difference spectra acquired before and after each irradiation. Irradiation conditions are: 3279~\CM, duration 15~min, average laser power 51~mW; 2222~\CM, 15~min, 33~mW; 1666~\CM, 15~min, 285~mW; 847~\CM, 15~min, 417~mW. Differences are: [spectrum after irradiation - spectrum before irradiation]. Color-coded dashed vertical lines highlight from ``blue to red'' all irradiations at 3279, 2222, 1666, and 835~\CM. Spectra are offset for clarity. The asterisks indicate pollution from the heating of the cryogenic kapton tape. At 1666 and 847~\CM, the laser power is 285 and 417~mW, respectively. }\label{fig:BulkIR4}
\end{figure}

Figure~\ref{fig:BulkIR4} presents FTIR difference spectra acquired before and after individual IRFEL irradiations on the stretching (3279~\CM), combination (2222~\CM), bending (1666~\CM), and libration (847~\CM) modes. IRFEL irradiations of pASW at 18~K led to spectral changes that are similar to a first approximation, as previously observed for thinner pASW samples \cite{Noble20}. The difference spectra in Figure~\ref{fig:BulkIR4} display systematically the same behaviour, \textit{i.e.} overall, all the bands decrease except that at $\approx$ 930~\CM. However, the relative intensities of band profile changes and their peak positions are IRFEL irradiation dependent, with the largest variations seen around the stretching mode of water ice. If we consider an average H-bond strength in the ice of $\approx$ 1200~\CM, as derived by Nissan\cite{Nissan56}, we can roughly assume that irradiations above this theoretical threshold may be energetic enough to induce ice changes via H-bond breaking. We can further put this argument into context by considering the wealth of literature data on vibrational energy transfer between hydrogen-bonded molecular systems other than water. As reported by Dlott \textit{et al.}\cite{Dlott02}, in the case of non-cyclic alcohols in solution at room temperature, vibrational energy transfer is mainly mechanical and can be assisted by anharmonic coupling with the solvent (the so called "through bond transfer"). In the case of 1-propanol, the thermal bath will `provide' about 200~\cm energy to help the last relaxation step between the last -CH$_2$- and the terminal -CH$_3$ group, higher in energy by about 300~\CM. In the case of pASW at 18~K and irradiated in the libration mode at 847~\CM, we can rule out this scenario because upon irradiation at 847~\CM the thermal pASW bath will not help to reach 1200~\cm necessary to theoretically trigger ice changes via the breaking of hydrogen bonds. In this case, water molecules must reorganise without breaking the hydrogen bonding network or gain energy from the formation of new H-bonds in the ice (particularly at pore surfaces). Since the macropulse energy of FELIX-2 is around 835~mW at 847~\CM, based upon these purely energetic arguments transposed from other systems one could suspect that a two-photon absorption process is potentially causing changes seen in the FTIR difference spectra. However, in our recent work\cite{Noble20} on thinner pASW and Ic ices, we found a linear relationship between laser power and FTIR spectral changes in the ice by varying the laser power at the libration frequency, providing clear indication of a one-photon process. Further studies are needed across the full MIR spectral range to time-resolve IRFEL effects on ice materials in order to resolve this issue.

As can be seen in Figure~\ref{fig:BulkIR4}, we observed photo-induced effects depending on the irradiation frequency upon IRFEL exposure of the pASW ice sample, \textit{i.e.} vibrational selectivity, in agreement with what we observed upon irradiating the dangling modes in previous works\cite{Noble14A,Noble14B,Coussan15} and upon irradiating the same vibrational modes in a thinner pASW sample \cite{Noble20}. However, we should point out that differences between the IRFEL irradiations presented in Figure~\ref{fig:BulkIR4} are minor compared to the overall trend that seems to be reproduced regardless of the irradiation frequency used, that is overall negative peaks in the difference spectra except at the libration mode. Compared to our results on thinner pASW\cite{Noble20}, we rarely observe fully positive peaks in the 3~$\mu$m spectral region in the difference spectra upon IRFEL irradiation. This can be due to the fact that desorption is more favorable for thicker layers of ice and dominates the spectral changes observed. However, global heating of the ice does not fully reproduce the effects seen upon IRFEL irradiation, because IR irradiation seems to be more restricted and selective\cite{Noble14A,Noble14B,Coussan15,Mihisan05,Focsa03,Krasnopoler98}. A more systematic and in depth study of such phenomena will be the focus of future work. In this study, we can discriminate two types of pASW behaviour upon IRFEL selective irradiation: irradiation of the surface modes, leading to the photoinduced rearrangement of these oscillators; and irradiation on bulk modes, leading to photoinduced, on-resonance local thermal effects occurring after vibrational energy dissipation.   This is in agreement with Noble \textit{et al.}\cite{Noble20}, where we observe similar effects in thinner pASW samples.

\emph{Stephane: It would be great to add a comment about the fact that the final, saturated effect from btr or from rtb is the same, regardless of the different FELIX setting 5vs10 Hz, strengthening the argument that individual oscillator classes respond in the same way, and the overall effect is not time-dependnet and not two photon.}

In a previous work, we have already described the inhomogeneity of dangling bonds at the pASW surface\cite{Coussan15}. Our work presented here confirms some degree of inhomogeneity in the bulk modes.
As mentioned by Schober \textit{et al.}\cite{Schober00} in their X-ray Scattering (IXS) experiments on different water ice types, low density amorphous ice shows sharp phonon-like excitations as compared to the much broader features that are observed in other amorphous system. The authors ascribe  is to low local disorder in LDA, going so far as to call it `crystal-like' behavior. High density amorphous ice (HDA) also shows narrow features but to a lesser extend. The authors deduce from their data highly intact hydrogen bond networks both in LDA and to some lesser degree in HDA.
This is precisely what we illustrate here studying the vibrational relaxation properties of pASW, our results are consistent with dissipation of the vibrational excitation in the local environment through the locally structured hydrogen bond network. Amorphous water seems to present on a short scale range (nm) a kind of local order.

\section{Conclusions and Perspectives}
In this work, we report the irradiation of pASW from the stretching mode to the libration mode by means of an IRFEL. We show that IR photons induce modification of the pASW structure, a phenomenon which is wavelength dependent.
The effects can be divided into two classes: surface irradiations on the dangling modes, which induce a rearrangement of these oscillators; and bulk irradiations, which lead to a local vibrational excitation of the ice. Surface irradiation effects can be explained in terms of competition between the conversion of vibrational energy by surface molecules into rearrangement or dissipation. Bulk irradiation effects cause similar global spectral changes in the ice regardless of whether IRFEL irradiation is carried out at the stretching, combination, bending, or libration modes. Local heating remains limited in terms of irradiated volume of the sample compared to a global classical thermal heating. Upon IRFEL irradiation we do not observe a clear sign of crystallization. This is likely due to the fact that several different mechanisms are at play at the same time. In fact, desorption is likely the reason why we only see negative peaks in the difference spectra, apart from the low signal-to-noise libration mode.
We also report the presence of frequency dependent changes in the ice, especially visible in the relative band intensity and peak position of changes at the stretching mode of pASW in FTIR difference spectra. A similar frequency dependence was observed in our previous work on thinner samples of pASW. Our results show proof of energy dissipation pathways occurring in a material with some degree of local organization. Future studies will include the quantification of such effects by systematically investigating ice thickness,  ice morphology, and ice composition. Ubiquitous in the Universe, pASW is found pure and mixed with molecules such as N$_2$, O$_2$, CH$_4$, and up to larger species like polycyclic aromatic hydrocarbons (PAHs). It is thus of a prior importance to understand the interact of such metastable material with IR radiation with myriad applications, from astronomy, astrochemistry, and astrobiology to climate change.

\section*{Credit authorship contribution statement}
S. Ioppolo initiated and managed the project at HFML-FELIX. S. Coussan wrote the manuscript with assistance from S. Ioppolo, J.A. Noble and H. Cuppen. S. Coussan, J.A. Noble, and S. Ioppolo performed all laboratory experiments. All authors contributed to data interpretation and commented on the paper.

\section*{Declaration of Competing Interest}
The authors declare that they have no known competing financial interests or personal relationships that could have appeared to influence the work reported in this paper.

\section*{Acknowledgements}
The authors thank the HFML-FELIX team for their experimental assistance and scientific support. The LISA end station is designed, constructed, and managed at the HFML-FELIX by the group of S. Ioppolo. This work was supported by the Royal Society University Research Fellowships Renewals 2019 (URF/R/191018); the Royal Society University Research Fellowship (UF130409); the Royal Society Research Fellow Enhancement Award (RGF/EA/180306); and the Royal Society Research Grant (RSG/R1/180418). Travel support was granted by the UK Engineering and Physical Sciences Research Council (UK EPSRC: Grant No. EP/R007926/1 entitled, FLUENCE: Felix Light for the UK: Exploiting Novel Characteristics and Expertise), and short term scientific missions (COST Actions CM1401 and TD1308).

\newpage

\section*{Authors information}
*Corresponding authors : St\'{e}phane Coussan, stephane.coussan@univ-amu.fr; Sergio Ioppolo, s.ioppolo@qmul.ac.uk

\end{document}